\documentclass[aps,pra,twocolumn,superscriptaddress,
showpacs,showkeys,amsmath,amssymb]{revtex4-1}
\usepackage{graphicx}
\usepackage{dcolumn}
\usepackage{bm}
\usepackage{textcomp}
\usepackage{stmaryrd}
\usepackage{datetime} 
\usepackage{wasysym}
\usepackage{marvosym} 
\usepackage{pifont} 
\usepackage{color}
\def\1{\mathchoice{\rm 1\mskip-4.2mu l}{\rm 1\mskip-4.2mu l}{\rm
        1\mskip-4.6mu l}{\rm 1\mskip-5.2mu l}}

\newcommand{\sio}{\mbox{$\rm SiO_2$}} 

\usepackage[dvips]{hyperref}
\begin{document}

\title{Diffraction of slow neutrons by holographic SiO\texorpdfstring{$_2$}{2} nanoparticle-polymer composite gratings}

\author{J. Klepp}
\email[]{juergen.klepp@univie.ac.at}
\homepage[]{http://fun.univie.ac.at} 
\affiliation{University of Vienna, 
Faculty of Physics, 1090 Wien, Austria}
\author{C. Pruner}
\affiliation{University of Salzburg, 
Department of Materials Science and Physics, 5020 Salzburg, Austria}
\author{Y. Tomita}
\affiliation{University of Electro-Communications, 
Department of Engineering Science, 1-5-1 Chofugaoka, Chofu, Tokyo 182, Japan}
\author{C. Plonka-Spehr} 
\affiliation{University of Mainz, Institute for Nuclear Chemistry, 55128 Mainz, Germany}
\author{P. Geltenbort}
\author{S. Ivanov}
\author{G. Manzin}
\affiliation{Institut Laue Langevin, 
Bo\^{i}te Postale 156, F-38042 Grenoble Cedex 9, France}
\author{K.H. Andersen} 
\affiliation{Institut Laue Langevin, 
Bo\^{i}te Postale 156, F-38042 Grenoble Cedex 9, France}
\affiliation{European Spallation Source, P.O. Box 176, 22100 Lund, Sweden}
\author{J. Kohlbrecher} 
\affiliation{Laboratory for Neutron Scattering, Paul Scherrer Institut, 5232 Villigen PSI, Switzerland}
\author{M. A. Ellabban}
\affiliation{Taibah University, Faculty of Science, Physics Department, 30002 Madinah, Saudi Arabia} 
\altaffiliation[On leave from ]{Tanta
University, Faculty of Science, Physics Department, Tanta 31527, Egypt}
\author{M. Fally}
\affiliation{University of Vienna, 
Faculty of Physics, Boltzmanngasse 5, 1090 Vienna, Austria}
\date{\today} 

\begin{abstract}
Diffraction experiments with holographic gratings recorded in SiO$_2$ nanoparticle-polymer composites have been carried out with slow neutrons. The influence of parameters such as nanoparticle concentration, grating thickness and grating spacing on the neutron-optical properties of such materials has been tested. Decay of the grating structure along the sample depth due to disturbance of the recording process becomes an issue at grating thicknesses of about 100 microns and larger. This limits the achievable diffraction efficiency for neutrons. As a solution to this problem, the Pendell\"{o}sung interference effect in holographic gratings has been exploited to reach a diffraction efficiency of 83\% for very cold neutrons.
\end{abstract}
\pacs{82.35.Np, 42.40.Eq, 03.75.Be}
\keywords{Nanoparticles in polymers,
Holographic optical elements, Holographic gratings, Neutron optics}
\maketitle
\section{Introduction\label{intro}}
Neutron optics \cite{Sears-89} is governed by the one-particle Schr\"{o}dinger equation 
that contains the neutron-optical potential -- equivalent to 
the neutron refractive index $n$ for the incident wavelength $\lambda$.
As in light optics, a basic example is diffraction by a one-dimensional sinusoidal grating
characterized by the periodically modulated refractive index 
\begin{eqnarray}\label{eq:grating}
n(x)=\overline n+\Delta n \cos\left(\frac{2\pi}{\Lambda} x\right).
\end{eqnarray}
Here, $\overline n$, $\Delta n$ and $\Lambda$ are the homogeneous refractive index of the grating material, the refractive-index modulation and the grating spacing, respectively. 
Depending on its diffraction efficiency -- or reflectivity -- diffraction gratings for neutrons can, in principle, be used for various purposes such as wavelength filters, guides, mirrors or beam-splitters for matter-wave interferometry with neutrons.

Mach-Zehnder-type perfect-crystal neutron interferometers for thermal neutrons \cite{Rauch-pla74} have played an important role in investigations of fundamental physics \cite{Rauch-00,HasegawaPRL2001,
RauchNature2002,
HasegawaNature2003,PushinPRL2008,HuberPRL2009,
BartosikPRL2009,HasegawaPRA2010}. 
The Mach-Zehnder geometry for neutrons was also implemented using artificial structures, such as Ni gratings in reflection geometry combined with mirrors \cite{IoffePLA1985}, or thin transmission phase gratings -- sputter-etched in quartz glass -- for very cold neutron interferometry \cite{Gruber-pla89,VanDerZouwNIMA2000}.
Further, multilayer-mirrors have been employed for cold neutron interferometry \cite{Funahashi-pra96}.
Phase and absorption gratings fabricated by photolithography have been used for neutron phase-imaging and tomography \cite{PfeifferPRL2006}.

Artificial grating structures can also be produced by exploiting the light-induced change of the refractive index for light in nonlinear optical materials -- the photorefractive effect. The phenomenon has been studied intensively since its discovery in 1966 \cite{AshkinAPL1966}. It was realized only in 1990 \cite{Rupp-prl90,MatullEPL1991} that its analogue for neutrons exists: The light-induced change of the refractive index for neutrons in materials -- the photo-neutron-refractive effect -- has been exploited for producing diffraction gratings for neutrons by structuring suitable recording materials using holography \cite{Fally-apb02}. 
Upon illumination with a spatial light pattern, a neutron refractive-index change occurs that is proportional to the change of the coherent scattering length density. In the case that only one isotope is present, the latter quantity is the product of the coherent scattering length $b_c$ of the isotope and its corresponding number density $\rho$. 
In holography, signal and reference light beams
are superposed at the position of a recording material. If the superposition results in a sinusoidal light intensity pattern -- modulating $\rho$ in $x$-direction due to a spatially inhomogeneous photopolymerization process, say -- neutron diffraction gratings are recorded. For neutrons, the corresponding refractive-index modulation reads as
$\Delta n=\lambda^2\cdot b_c\Delta\rho/(2\pi)$. 
The quantity $b_c\Delta\rho$ is referred to as coherent scattering length density modulation.
Since diffraction of neutrons by such gratings essentially constitutes a readout of the hologram, it is convenient to adopt Kogelnik's two-wave coupling theory for Bragg diffraction of light by holographic volume phase gratings \cite{KogelnikBellSysJ1969} to write the diffraction efficiency for neutrons in the symmetric Laue-case (transmission geometry) as 
\begin{eqnarray}\label{eq:Kogelnik}
\eta=\nu^2\frac{\sin^2\sqrt{\nu^2+\xi^2}}{\nu^2+\xi^2},
\end{eqnarray} 
with 
\begin{eqnarray}\label{eq:nu}
\nu=\frac{\lambda\,d_0\, b_c\Delta\rho}{2\cos\theta}\,\,\mbox{and}\,\,\xi=\frac{\pi d_0(\theta_B-\theta)}{\Lambda}.
\end{eqnarray} 
Here, $d_0$, $\theta$ and $\theta_B$ are the thickness of the grating, the angle of incidence and the Bragg angle (as defined by $\lambda=2\Lambda\sin\theta_B$), respectively. Note that Eq.\,(\ref{eq:Kogelnik}) is equivalent to the expression for the reflectivity of thick crystals in Laue-geometry derived from dynamical diffraction theory \cite{Sears-89}, as was discussed in Ref.\,\cite{KleppNIMA2011}.   
Our goal is to adjust the tunable parameters $d_0,\,b_c,\,\Delta\rho$ and $\Lambda$ so that suitable peak-values of $\eta$ ($\eta\equiv\eta_P$ for $\theta=\theta_B$) and widths of the reflectivity- or rocking curve desired for a particular application are reached. For instance, three of these gratings in Laue-geometry can be used for setting up a Mach-Zehnder type neutron interferometer \cite{Schellhorn-phb97,Pruner-nima06}, by tuning the parameters so that $\eta_P=50$\% (beam splitters) for the first and the last grating and $\eta_P=100$\% (mirror) for the second grating.   

Various material classes have been investigated for holographic production of neutron-diffractive elements. 
Diffraction gratings recorded in deuterated polymethylmethacrylate \cite{Schellhorn-phb97,
Pruner-nima06} exhibit high reflectivity, but the recorded patterns are not stable and -- due to continuing polymerization processes -- change their diffraction properties within weeks. Stability on a time scale of years is highly desirable for neutron-optical devices. Holographic polymer-dispersed liquid crystals \cite{FallyPRL2006} have also been considered. The best reflectivity that has been reached so far is about 3\% at a wavelength of 2 nm. 
Recently, another material class was found suitable for producing neutron-optical devices: nanoparticle-polymer composites. Inorganic nanoparticles embedded in a photopolymer matrix have been investigated intensively \cite{SuzukiAPL2002,SuzukiApplOpt2004,
TomitaOptLett2005,
SuzukiOptExp2006,
SakhnoNanotech2007,
NakamuraJOptA2009,SakhnoJOptA2009}.
In these materials, the refractive-index modulation $b_c\Delta\rho$ can be tuned by the species of nanoparticles and their concentration. 
Furthermore, including nanoparticles in the polymer matrix increases the mechanical stability, i.e. shrinkage -- typical for polymerization processes -- is strongly reduced \cite{SuzukiAPL2002}. 
Moreover, it was shown that relatively thick gratings can be produced without serious problems that might be expected to occur because of light-scattering during the recording process \cite{SuzukiApplOpt2007}. Only recently, the feasibility of a beam splitter for neutrons at a wavelength of 2 nm has been demonstrated using holographic diffraction gratings recorded in such materials \cite{FallyPRL2010}.

In this article we present a detailed description and further results of neutron diffraction experiments with holographic \sio\,nanoparticle-polymer gratings that have led to the recent achievements \cite{FallyPRL2010}. Various gratings differing in nanoparticle concentration, grating thickness $d_0$ and grating spacing $\Lambda$ have been tested with neutron wavelengths in the range of 1.7\,nm to 3.76\,nm. The Pendell\"{o}sung interference effect \cite{Shull-prl68,Sears-89} allows to further increase the neutron reflectivity by tilting gratings of limited thickness around an axis parallel to the grating vector \cite{SomenkovSolStComm1978}. A reflectivity of up to 83\% has been reached for a neutron wavelength of 3.76\,nm without significant loss of intensity to higher diffraction orders. 

\section{Grating preparation}\label{sec:GratingPrep}
The \sio\,nanoparticles used for the present investigation have an average core diameter of about 13 nm. They are produced by liquid-phase synthesis and dissolved in a methyl isobutyl ketone (toluene) solution. The \sio\,sol is dispersed to methacrylate monomer whose
refractive index for light is 1.59 in the solid phase at 589 nm. At this wavelength the refractive index of the surface-treated \sio\, nanoparticles is 1.46\,.

As radical photoinitiator, 1wt.\% titanocene (Irgacure784, Ciba) is added to enable the monomer to photo-polymerize at wavelengths shorter than 550 nm. The mixed syrup is cast on a glass plate and is dried. Spacers are arranged around the sample before it is covered with another glass plate to obtain film samples ready for being structured by a spatial light-intensity pattern.
At this stage of the preparation, the photoinitiator, the monomer and the nanoparticles are homogeneously distributed in the sample material. 

Next, two expanded, mutually coherent and $s$-polarized laser beams of equal intensities and a wavelength of 532 nm are superposed to create a spatial sinusoidal light-intensity pattern at the sample position. In the present case, the beam diameter was about 1 cm.
Via the photoinitiator, the pattern induces polymerization in the bright sample regions, a process that consumes monomers that migrate from dark to bright regions \cite{TomitaOptLett2005}. As a consequence of the growing monomer-concentration gradient, 
nanoparticles diffuse from bright to dark regions, resulting in an approximately sinusoidal nanoparticle-concentration 
pattern. This forms the hologram -- the diffraction grating. Subsequent homogeneous illumination ensures that the material is fully polymerized so that the nanoparticle density-modulation remains stable.
Further details on the sample preparation technique can be found in Ref. \cite{SuzukiApplOpt2004}. 

Samples were prepared so that two of a particular pair differ from each other only in a single parameter -- concentration of nanoparticles (20 vol\% or 34 vol\%), thickness (from about 50 to 240 microns) or grating spacing (0.5 or 1 micron) -- to separate its corresponding effect on the reflectivity.

\begin{figure}
    \scalebox{0.4}
{\includegraphics {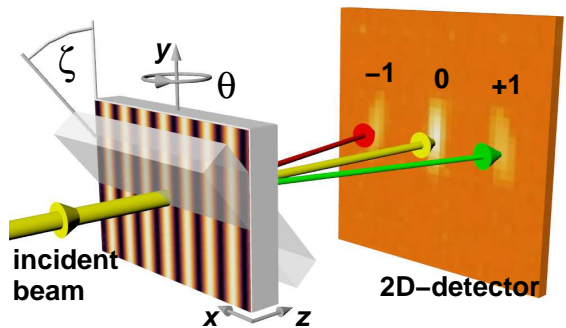}}
    \caption{Sketch illustrating the measurement principle.}
    \label{fig1}
\end{figure}

\section{Experiments}\label{sec:Experiments}

Neutron diffraction experiments were carried out at the instruments SANS I of the SINQ spallation source at the Paul Scherrer Institut (PSI) in Villigen, Switzerland and PF2 of the Institut Laue-Langevin (ILL) in Grenoble, France.
The measurement principle is sketched in Fig.\,\ref{fig1}. As mentioned above, the gratings were analyzed in Laue-geometry. Tilted to the angle $\zeta$ around an axis parallel to the grating vector (in order to adjust the effective thickness), the incident angle $\theta$ was varied to measure rocking curves in the vicinity of $\theta_B$. 

At SANS I of SINQ, the wavelength distribution of the neutrons $\Delta\lambda/\lambda$, as incident from a velocity selector, is typically 10\%.
The collimation slit-width and distance were chosen so that the typical beam divergence was better than 0.06\textdegree. 
Measurements were carried out with various samples for 1.7 nm and 2 nm incident neutron wavelengths. 
At PF2 of the ILL, very cold neutrons with a broad wavelength distribution in the range of about 3 to 6 nm are available \cite{YamadaNIMA2011}. A neutron mirror -- a Ti/Ni-mirror in this case -- is necessary to redirect the incident neutron beam to make use of the full 
collimation length and to obtain a narrower wavelength distribution. Here, the divergence of the beam was better than 0.1\textdegree. 
In both setups, 2D-detectors were used.
Depending on available flux, collimation and specific sample, typical measurement times per setting of the incident angle $\theta$ range from a couple of minutes to $1.5$ hours.
For a more complete description of the beamlines see, e.g.,
Refs.\,\cite{KohlbrecherJApplCryst2000} 
and \cite{YellowBookILL2008}. 

Since the grating spacing $\Lambda$ is large compared to the wavelength of the incident neutrons, $\theta_B\sim\lambda/(2\Lambda)$ is of the order of 0.1\textdegree. Depending on the spatial resolution of the detector system, that is of the order of mm, some meters distance have to be maintained between the grating and the detector. 

We define the (relative) diffraction efficiency or reflectivity of the $\pm 1^{\mbox{\scriptsize{st}}}$ diffraction order as 
\begin{eqnarray}\label{eq:diffrEff}
\eta_{\pm 1}=\frac{I_{\pm 1}}{I_0+I_{+1}+I_{-1}}, 
\end{eqnarray} 
where $I_0$ and $I_{\pm 1}$ are the measured intensities of the $0^{\mbox{\scriptsize{th}}}$ and $\pm 1^{\mbox{\scriptsize{st}}}$ diffraction orders, respectively. For each incident angle, the sum over all 2D-detector pixels in each separated spot (see Fig.\,\ref{fig1}) -- associated to one diffraction order -- was calculated and the resulting intensities (corrected for background) plugged into Eq.\,(\ref{eq:diffrEff}).
Second order diffraction was neglected in the analysis as well as in the data plots of Sec.\,\ref{sec:MeasResults} wherever its contribution is marginal, which was the case for almost all the measurements. 

Note that one of the assumptions to arrive at Eq.\,(\ref{eq:Kogelnik}) is two-wave coupling, i.e. only two waves are propagating through the sample at a particular incident angle $\theta$. This is not always the case, as will be seen from the overlapping $\pm 1^{\mbox{\scriptsize{st}}}$ diffraction orders in some of the data plots. However, as long as diffraction efficiencies for the $\pm 1^{\mbox{\scriptsize{st}}}$ diffraction orders are rather small, one can neglect the additional peak for either case and use a two-wave coupling theory for the analysis of the rocking curves. 

\section{Measurement results}
\label{sec:MeasResults} 

\begin{figure*}
    \scalebox{0.52}
{\includegraphics {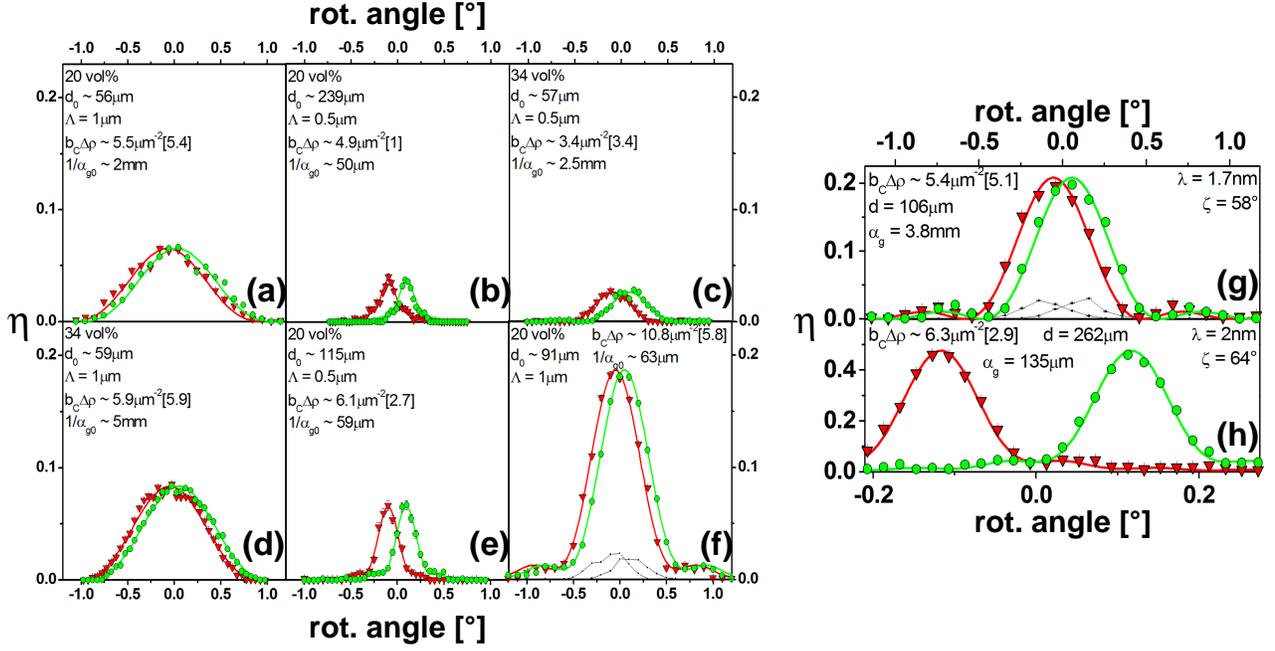}}
    \caption{Cold neutron data (SANS I): (a) -- (f) Rocking curves of various gratings measured at $\lambda= 1.7$nm. 
(g) -- (h) Rocking curves of two samples corresponding to plots shown in Figs.\,\ref{fig2}(a) and \ref{fig2}(e) tilted to 58\textdegree\,and 64\textdegree, respectively. (\textcolor{red}{\ding{116}}
$\dots -1^{\mbox{\scriptsize{st}}}$ order, \textcolor{green}{\ding{108}}$\dots +1^{\mbox{\scriptsize{st}}}$ order)}
    \label{fig2}
\end{figure*}

In Figs.\,\ref{fig2} and \ref{fig3}(a), rocking curves for various samples are plotted. 
For each rocking curve, least-squares fits of the Kogelnik theory, additionally assuming exponential decay of $\Delta n$ along the $z$-direction, i.e. $\Delta n(z)=\Delta ne^{-\alpha_{g_0} z}$ were made (solid lines). Here, $z$ is the sample depth -- the distance as measured from the sample front surface in direction of the grating thickness. The inverse of the parameter $\alpha_{g_0}$ is the sample depth at which $\Delta n$ has decreased to $1/e$ of its value at the front surface. Such a behavior was described by Uchida \cite{UchidaJOptSocAm1973}. It can, for instance, occur because of holographic light-scattering \cite{FallyPRB2000,EllabbanAPL2005,
ImlauPRB2006} that smears out the recording light-intensity pattern within thick nanoparticle-polymer gratings \cite{SuzukiApplOpt2007}. 
In each plot of Figs.\,\ref{fig2} and \ref{fig3}(a), the fitted parameters for the corresponding grating are given. 
Fixed parameters are given by relations using `=', while we use `$\sim$' for fitted parameters in the plots throughout the rest of the paper.
Estimations of relative errors for the fitted parameters are typically some percent. For very large values of $1/\alpha_{g_0}$ in comparison to $d_0$, the data are well-described by the Kogelnik theory alone. 
Thus, the exact value of $\alpha_{g_0}$ is insignificant.
We write $\langle b_c\Delta\rho\rangle$ for the coherent scattering length density modulation averaged over the thickness $d_0$ as
\begin{eqnarray}\label{eq:average}
\langle b_c\Delta\rho\rangle=\frac{b_c\Delta\rho}{d_0}\int_0^{d_0}
e^{-\alpha_{g_0} z}dz=\frac{b_c\Delta\rho}{\alpha_{g_0} d_0}\left(1-e^{-\alpha_{g_0} d_0}\right).
\end{eqnarray} 
In Figs.\,\ref{fig2} and \ref{fig3}(a), $\langle b_c\Delta\rho\rangle$ is given in units of $\mu$m$^{-2}$ by the value in brackets next to the value for $b_c\Delta\rho$.   

\subsection{Nanoparticle concentration}\label{subsec:nanopConcentr}
From previous studies in light optics, it is expected that one can increase $\Delta\rho$ to a certain degree by increasing the concentration of nanoparticles in the sample material \cite{SuzukiApplOpt2004,SuzukiOptExp2006}. In Figs.\,\ref{fig2}(a) and \ref{fig2}(d), the rocking curves for gratings with essentially equal sample parameters, except for nanoparticle concentrations of 20 vol\% and 34 vol\%, respectively, are shown. 
For light optics, a value of 34 vol\% of SiO$_2$ nanoparticles yielded the highest values of $\Delta n$ for gratings thinner than $100\mu$m \cite{SuzukiApplOpt2004}.
By comparison of Figs.\,\ref{fig2}(a) and \ref{fig2}(d), one can see that -- although showing a slight increase of $b_c\Delta\rho$ for 34 vol\% -- the estimates of $b_c\Delta\rho$ as obtained from the fits do not differ substantially. 
From the light optical investigations, the possibility of a rather flat or even non-monotonous dependency of $\Delta n$ on the nanoparticle concentration in the range between 20 vol\% and 34 vol\% -- with a maximum between the two concentrations, say -- cannot be ruled out. After reaching a maximum, $b_c\Delta\rho$ decreases for further increasing nanoparticle concentration because counter-diffusion of monomers and nanoparticles cannot be improved further by adding even more nanoparticles \cite{SuzukiOptExp2006}. Further experiments must be carried out to see if variation of the nanoparticle concentration still leaves the scope to further optimize $\Delta\rho$ by such means.
Note that in the cases shown in Figs.\,\ref{fig2}(a) and \ref{fig2}(d), $1/\alpha_{g_0}$ is of the order of several mm -- much larger than the grating thickness. As has also been confirmed by measurements with light (not shown), decay of $\Delta n$ is not a problem for rather thin \sio\,gratings ($d_0\simeq 60\mu$m).

\subsection{Grating thickness}\label{subsec:gratThickn}
Next, consider Figs.\,\ref{fig2}(b) and \ref{fig2}(e). The two gratings used here only differ in thickness. It is expected, that possible disturbances that lead to decay of $\Delta n$ along the sample depth are observed only in thicker samples \cite{SuzukiApplOpt2007}. As can be seen in the plots, this is not confirmed by the neutron measurements. The small value of $1/\alpha_{g_0}\!\sim\! 59\mu$m, i.e. that $\Delta n$ drops to $\Delta n/e$ at about half the sample thickness of only $115\mu$m, is an obstacle for tuning $\eta_P$ for a particular application, since it sets a rather low limit for $d_0$. This value of $1/\alpha_{g_0}$ has also been confirmed by hologram readout with light (not shown). 
The result may be attributed to the rather small grating spacing ($\Lambda=0.5\mu$m), which has been demonstrated to have negative effects on $\Delta n$ \cite{SuzukiAPL2002,
SuzukiApplOpt2004,SuzukiOptExp2006}.  
On the other hand, comparing rocking curves for gratings that again only differ in thickness but have $\Lambda=1\mu$m (see Figs.\,\ref{fig2}(a) and \ref{fig2}(f)), one can see that the problem of decaying $\Delta n$ already at smaller thicknesses persists: While decay is reduced dramatically for $d_0\!\sim\! 56\mu$m, it is clearly present for $d_0\!\sim\! 91\mu$m. 
As will be explained in Sec.\,\ref{sec:tilting}, a workaround to this problem is provided by making use of the Pendell\"{o}sung interference effect. Note also that in Fig.\,\ref{fig2}(f), small contributions of the $\pm 2^{\mbox{\scriptsize{nd}}}$ diffraction orders are plotted.  

\subsection{Grating spacing}\label{subsec:gratSpac}
As already mentioned above, it has been demonstrated using light that $\Delta n$ can be improved by increasing $\Lambda$ \cite{SuzukiAPL2002,
SuzukiApplOpt2004,SuzukiOptExp2006}. This behavior is also observed with neutrons comparing Figs.\,\ref{fig2}(e) and \ref{fig2}(f) for $d_0\simeq 100\mu$m (20 vol\%) or Figs.\,\ref{fig2}(c) and \ref{fig2}(d) for $d_0\simeq 60\mu$m (34 vol\%).
In both cases, $\langle b_c\Delta\rho\rangle$ roughly doubles for larger $\Lambda$ independent of the very different values for $1/\alpha_{g_0}$ of either pair. 
Again one can see that larger values of $\Lambda$ cannot prevent decay of $\Delta n$. It has been argued that formation of long monomer chains that reach far into dark regions could disturb the holographic recording at small grating spacings \cite{SuzukiOptExp2006}. Despite these facts, increasing the grating spacing is not a good option for producing neutron mirrors or beam splitters. It has two major drawbacks: \emph{i)} $\theta_B$ decreases and \emph{ii)} gratings with large $\Lambda$ do not produce two-wave regime diffraction (c.f. Fig.\,1 in \cite{FallyPRL2010}). 

\section{High reflectivity via the Pendell\"{o}sung effect}\label{sec:tilting} 
Shull \cite{Shull-prl68} demonstrated the Pendell\"{o}sung interference effect for neutrons, that is predicted by dynamical diffraction theory \cite{Sears-89}. 
The effect, that is also known in X-ray physics and light optics, arises due to the interfering components of the wave function within crystals or -- more general -- periodic structures. 
It results in an oscillation of the neutron flux between reflected and transmitted beams. 
The relevance of the Pendell\"{o}sung effect for neutron diffraction in holographic gratings has been pointed out in Ref.\,\cite{KleppNIMA2011}. There, based on Kogelnik's theory, the expression for the oscillation of the first order diffraction efficiency at the Bragg-position was written as: 
\begin{eqnarray}\label{eq:PendellOsc}
\eta_P=\sin^2\left(\frac{\pi d}{\Delta_K}\right).
\end{eqnarray} 
Here, $\Delta_K$ is called Pendell\"{o}sung period or extinction length and is -- in our case -- given by 
\begin{eqnarray}\label{eq:PendellPer}
\Delta_K=\frac{2\pi\cos\theta_B}{\lambda\, b_c\Delta\rho}
\end{eqnarray} 
for holographic gratings. 
As becomes clear from the results shown in Fig.\,\ref{fig2}, $\Delta_K$ ranges from 0.6 to 3.7mm for holographic nanoparticle-polymer gratings at the used wavelengths. 
The oscillation of intensity between reflected and transmitted neutron beams has been observed by variation of the incident wavelength \cite{Shull-prl68} and by tilting the sample around an axis parallel to the grating vector, so that essentially its effective thickness $d$, appearing in Eq.\,(\ref{eq:PendellOsc}), is increased \cite{SomenkovSolStComm1978}. 

Also for the experiments described in the following, the reflectivity was enhanced by tilting (tilt angle $\zeta$, see Fig.\,\ref{fig1}).  
For holographic gratings, tilting actually brings about the solution to the problem posed by decay of $\Delta n$ in $z$-direction: Whenever increasing the grating thickness by simply using thicker spacers between the two glass plates is not an option due to problems resulting from unwanted light-scattering in the recording process, one can increase the effective thickness of a thin grating by tilting. The decay parameter $\alpha_{g_0}$ as well as the thickness scale with the tilt angle $\zeta$, i.e. $\alpha_g=\alpha_{g_0}\cos\zeta$ and $d=d_0/\cos\zeta$. Consequently, the exponent and the denominator on the right hand side of Eq.\,(\ref{eq:average}) and, therefore, $\langle b_c\Delta\rho\rangle$ are kept constant. Thus, one should be able to raise the grating thickness, and thereby $\eta_P$, beyond the limit set by decay without much influencing the other parameters.

As a proof of principle, two of the sample gratings, namely the ones with rocking curves plotted in Figs.\,\ref{fig2}(a) and \ref{fig2}(e) were tilted to $\zeta=58$\textdegree\, and $\zeta=64$\textdegree, respectively. The results are shown in Figs.\,\ref{fig2}(g) and \ref{fig2}(h). 
As expected, the reflectivity compared to the measurements at $\zeta=0$\textdegree\, is increased according to Eq.\,(\ref{eq:PendellOsc}). With effective thickness $d$ and decay parameter $\alpha_{g}$ calculated from the corresponding values at $\zeta=0$\textdegree\,and the actual tilting angle $\zeta$, the results for $\langle b_c\Delta\rho\rangle$ as obtained from the fits (solid lines) are in reasonable accordance with the values given in Figs.\,\ref{fig2}(a) and \ref{fig2}(e). 
In the second case an incident wavelength of $\lambda=2$nm was used to further increase $\Delta n$ and demonstrate the feasibility of a beam splitter for cold neutrons. These data have already been presented in Ref.\,\cite{FallyPRL2010}.

Since our aim is the design of neutron-optical elements for long wavelengths, we chose one of the samples to be tested with very cold neutrons.

\begin{figure*}
    \scalebox{0.55}
{\includegraphics {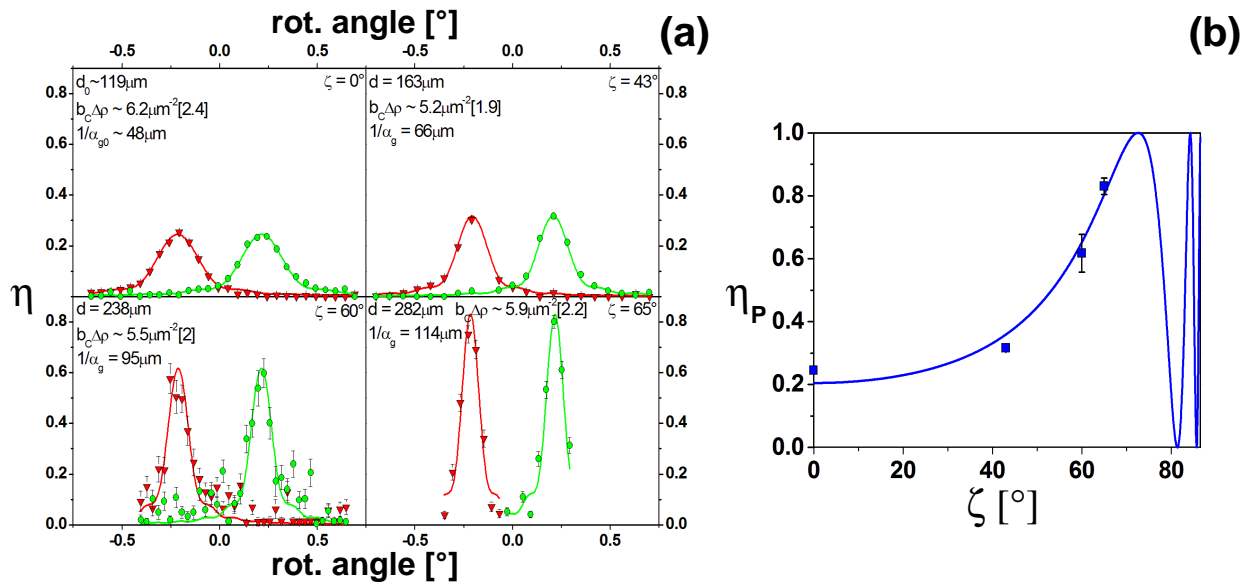}}
    \caption{Very cold neutron data (PF2): (a) Rocking curves of sample grating with 20 vol\%, $\Lambda=0.5\mu$m (see also Figs.\,\ref{fig2}(e) and \ref{fig2}(h)) for various tilt angles (\textcolor{red}{\ding{116}}
$\dots -1^{\mbox{\scriptsize{st}}}$ order, \textcolor{green}{\ding{108}}$\dots +1^{\mbox{\scriptsize{st}}}$ order).
(b) Peak reflectivity $\eta_P$ (\textcolor{blue}{\ding{110}}) versus $\zeta$ as determined from the rocking curves on the left compared with theory (solid line). 
}
    \label{fig3}
\end{figure*}

In Fig.\,\ref{fig3}(a), rocking curves for the same grating as used in Figs.\,\ref{fig2}(e) and \ref{fig2}(h) at $\lambda= 3.76$nm and four different angles $\zeta$ are shown.
The values for $d$ and $\alpha_g$ given in the plots were calculated from the fitting result $d_0\sim 119\mu$m (for $\zeta=0$\textdegree) and the actual tilt angles $\zeta$. One can see that $\langle b_c\Delta\rho\rangle$ remains approximately constant, as suggested above in this section. The highest peak-reflectivity reached is about 83\%. In Fig.\,\ref{fig3}(b), $\eta_P$ as extracted from the fits in Fig.\,\ref{fig3}(a) is plotted versus the tilt angle $\zeta$ together with a theory curve (solid line) according to Eq.\,(\ref{eq:PendellOsc}). The qualitative agreement is obvious. Even higher reflectivity or demonstration of a full Pendell\"{o}sung oscillation could not be reached due to the small sample height, that limits the measured intensities for large $\zeta$.

\section{Summary and Discussion}\label{sec:SummaryAndDisc}

Following recent achievements \cite{FallyPRL2010}, it has been demonstrated here that nanoparticle-polymer composites are a promising material class for realization of holographic gratings as versatile optical devices for slow neutron applications. The influences of material parameters such as nanoparticle concentration, grating spacing and thickness on neutron diffraction have been tested. Decay of the refractive-index modulation $\Delta n$ along the grating depth due to unwanted holographic light-scattering plays a role already at sample thicknesses of about 100 microns. However, it has been shown here, that the difficulty to obtain larger thicknesses can be overcome by exploiting the Pendell\"{o}sung interference effect. This was done by tilting the grating around an axis parallel to the grating vector to increase the effective thickness of the gratings \cite{SomenkovSolStComm1978}. With this method, high peak reflectivity of 83\% has been reached for a neutron wavelength of 3.76 nm. 
Tilting might also be useful for fine-tuning the reflectivity of holographic gratings in light optics applications. 
     
For large tilt angles, the sample area (1cm diameter) as seen from the incident direction appears very small in height. This strongly reduces the height of the observed diffraction spots and, therefore, the measured intensities. This problem can easily be overcome by expanding the recording laser beam to 2-3 cm as it has been done, e.g., in Ref.\,\cite{Pruner-nima06}. 
Also, the nanoparticle-polymer composites that were used for the present experiments contain hydrogen, which has large incoherent scattering and absorption cross-sections for neutrons. In future measurements, the signal-to-noise ratio can be improved by substitution of $^1$H by $^2$H, as it is done for polymethylmethacrylate gratings and with samples in many other neutron scattering experiments. Moreover, to further improve the gratings, quartz glass plates can be used as sample containers. For applications in light optics, the possibility of even removing the glass plates after recording of the holograms (`free standing films') has been considered \cite{VitaAPL2007}, which could also be an option for neutron diffraction.
 
The sensitivity needed to measure physical effects that become manifest in small lateral deflections could be provided by tilting of holographic gratings. For instance, mounting two suitably tilted gratings in a row, setting both to the Bragg-position and slightly varying the angle of incidence for the second grating with a prism results in 
a convolution of the individual rocking curves. This resulting rocking curve exhibits a very narrow so-called `central-peak' that has been shown to reach a width of the order $10^{-3}$ arcseconds \cite{BonsePL1979,RauchZPhysB1983} using perfect crystals and thermal neutrons. Implementing such a setup for very cold neutrons -- as it might be possible with holographic gratings -- could bring about new possibilities for fundamental physics experiments, such as the search for the neutron electric charge \cite{shullPR1967,BaumannPRD1988,PlonkaSpehrNIMA2010}.
\\
\begin{acknowledgments}
We thank R. Bebb for technical assistance at the ILL.
Financial support by the Austrian Science Fund (FWF): P-20265, by the Ministry of Education,
Culture, Sports, Science and Technology of Japan (Grant
No. 23656045) is greatly acknowledged. We are grateful for the hospitality of the ILL.
\end{acknowledgments}

\end{document}